# EXPERIMENTAL AND MODELING STUDY OF THE LOW-TEMPERATURE OXIDATION OF LARGE ALKANES

*Joffrey BIET, Mohammed Hichem HAKKA, Valérie WARTH, Pierre-Alexandre GLAUDE,*

*Frédérique BATTIN-LECLERC\**

Département de Chimie-Physique des Réactions, Nancy Université, CNRS, ENSIC, 1, rue Grandville,

BP 20451, 54001 NANCY Cedex - France

ABSTRACT

This paper presents an experimental and modeling study of the oxidation of large linear akanes (from $C_{10}$) representative from diesel fuel from low to intermediate temperature (550-1100 K) including the negative temperature coefficient (NTC) zone. The experimental study has been performed in a jet-stirred reactor at atmospheric pressure for n-decane and a n-decane/n-hexadecane blend. Detailed kinetic mechanisms have been developed using computer-aided generation (EXGAS) with improved rules for writing reactions of primary products. These mechanisms have allowed a correct simulation of the experimental results obtained. Data from the literature for the oxidation of n-decane, in a jet-stirred reactor at 10 bar and in shock tubes, and of n-dodecane in a pressurized flow reactor have also been correctly modeled. A considerable improvement of the prediction of the formation of products is obtained compared to our previous models. Flow rates and sensitivity analyses have been performed in order to better understand the influence of reactions of primary products. A modeling comparison between linear alkanes for $C_8$ to $C_{16}$ in terms of ignition delay times and the formation of light products is also discussed.

---

\* E-Mail : Frederique.Battin-Leclerc@ensic.inpl-nancy.fr, Tel : 33 3 83175125, Fax : 33 3 83378120

INTRODUCTION

If many detailed kinetic models are available for the oxidation of mixtures representative of gasolines, they are much less numerous in the case of diesel fuels because of their more complex composition. The constituents of diesel fuel contain from 10 to 20 carbon atoms and include about 30% (mass) of alkanes, the remaining part being mainly alkylcyclohexanes, alkyldecalines, alkylbenzenes and polycyclic naphtenoaromatic compounds (1). Many extensively validated models have been published for alkanes up to $C_8$ (e.g. (2-8)), but only a few for heavier compounds. The models related to light alkanes allow a satisfactory simulation of experimental results obtained in shock tubes, rapid compression machines, jet-stirred and flow reactors under a wide range of operating conditions, including pressures from 1 to 60 bar and temperatures range covering all the negative temperature coefficient (NTC) zone.

Table 1 summarizes the experimental results and the models proposed to simulate them which have been recently published for the low-temperature ignition and oxidation (below 900 K) of linear alkanes containing 10 or more carbon atoms. The first work related to the modeling of the low-temperature oxidation of alkanes from $C_{10}$ using a detailed mechanism was performed by Chevalier et al. (9-10) for n-decane and n-hexadecane. Simulations using their mechanism for n-decane lead to a good agreement with ignition delay times measured in a shock tube at 12 bar (11). Later Battin-Leclerc et al. (12) have proposed a detailed model for the low-temperature oxidation of n-decane with validation using data from a jet-stirred reactor (13) obtained at 10 bar in the temperatures range 550-1100 K covering all the NTC zone. This model has been up-dated by Buda et al. (7) to reproduce ignition data from a shock tube (11) with a correct agreement at 12 bar, which considerably deteriorated at 50 bar. Zhukov et al. (14) have used this mechanism to model their recent measurements of ignition delay times in a shock tube and shown a large overprediction at 80 bar. The team of Ranzi and Faravelli has proposed a semi-detailed model in the case of n-decane (15-16) with satisfactory validation for shock tube data at both 12 and 50 bar (11) and for jet-stirred reactor results between 550 and 1100 K (17). They have also proposed a model for n-dodecane (16) validated against species measurements made in a turbulent flow tube (18-19). Lastly Westbrook et al. (20) have proposed a detailed mechanism for n-alkanes from octane to n-hexadecane. Their model for the low-temperature oxidation of n-decane is able to well represent the shock tube data of Pfahl et al. (11) and Zhukov et al. (14) at 12 bar, but largely overpredicts the results obtained at 50 and 80 bar. These authors (20) have modeled the measurements of delay times of n-decane performed by Kumar et al. (21) using a rapid compression machine at 14.3 bar and attributed the observed faster results compared to both the shock tube experimental data (11, 14) and simulations to reaction during the compression stroke. A satisfactory agreement has been obtained for the oxidation of n-decane in a jet-stirred reactor (22) at temperature between 750 and 1200 K, above the NTC zone. The mechanism can also well predict the variations with temperature of CO concentrations in a turbulent flow tube (18). Experimental results have also been published for the oxidation of n-tetradecane in a flow reactor, but the fact that the fuel was injected as droplets prevents their use to validate simulations with a simple physical model (23).

## TABLE 1

The first objective of this paper is to present new experimental results for the low-temperature oxidation of large linear alkanes (n-decane and a n-decane/n-hexadecane blend) in a jet-stirred reactor at a lower pressure than that used by Dagaut et al. (13)(17). N-decane has been chosen because it is the most studied large alkane as shown in Table 1. N-hexadecane is a primary reference fuel for cetane rating for auto-ignition ability, with a cetane number of 100 ($\alpha$-methyl-naphthalene has a cetane number of 0). The interest of studying a mixture of these two compounds is to investigate the potential effects of the addition of a larger alkane on the oxidation of n-decane. The second purpose of this work is to describe the improvements made in kinetic models to consider reactions of primary products of fuel oxidation with more details for a better modeling of the low-temperature oxidation of large alkanes using for validation the experimental results described here, as well as those of the literature listed in Table 1. The global reactions of primary products previously considered to reproduce the reactivity of alkenes or oxygenate products lead to deteriorate results for larger alkanes. A third objective is to perform a modeling comparison between the reactivity of the alkanes from $C_7$ to $C_{16}$. The purpose of this comparison is to help to determine the more suitable model molecules to represent the large linear alkanes present in diesel fuel.

As this paper is interested with the chemistry involved during the low-temperature oxidation of alkanes, let us remind its main features as they are currently understood from the work of Pollard (24), Cox and Cole (25) and Walker and Morley (26). Figure 1 shows a simplified scheme of the main reactions, which are now usually admitted to model the oxidation of an alkane (RH). At low temperature, i.e. below 900 K, the alkyl radicals R• are produced by H-abstractions mostly by •OH radicals and in a lesser extent by •HO$_2$, •H, •CH$_3$ and CH$_3$OO• radicals. They add thereafter to O$_2$ and yield peroxy radicals ROO•. Some consecutive steps will then lead to the formation of hydroperoxides, which are degenerate branching agents explaining the high reactivity of the alkanes at low temperature. Peroxy radicals ROO• will first isomerize to hydroperoxyalkyl radicals •QOOH. This first isomerization has a very large influence on the low-temperature oxidation chemistry; it occurs via an internal H-abstraction going through a cyclic transition state.

**FIGURE 1**

The •QOOH radicals can thereafter add on O$_2$ and yield some •OOQOOH radicals that react through a second internal isomerization producing •U(OOH)$_2$ radicals. The rapid unimolecular decomposition of •U(OOH)$_2$ radicals is favored over further bimolecular O$_2$ addition. This decomposition pathway eventually produces three radicals and is the source of chain branching occuring at low temperatures:

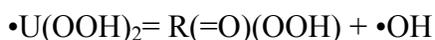

•U(OOH)$_2$ = R(=O)(OOH) + •OH

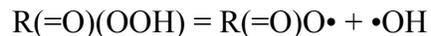

R(=O)(OOH) = R(=O)O• + •OH

In this scheme, a •OH radicals reacting by H-abstraction with the initial alkane allows the formation of three free radicals, two •OH and a R(=O)O• radical. This auto-accelerating process strongly promotes the oxidation of alkanes below 700 K.

When the temperature increases, the equilibrium R• + $O_2$ = ROO• begins to be displaced back to the reactants and the formation of ROO• radicals is less favoured. The larger concentration of R• radicals increases the rate of the strongly inhibiting oxidation reaction R• + $O_2$ = •$HO_2$ + conjugated olefin. This reaction is equivalent to a termination step since •$HO_2$ radicals react mostly by the disproportionation reaction: 2 •$HO_2$ = $H_2O_2$ + $O_2$. Furthermore, unimolecular reaction channels from •QOOH radicals, which have a high activation energy and involve fragmentation into various products, become faster and compete with $O_2$ addition. •QOOH radicals can decompose into cyclic ethers or ketines and •OH radicals or by β-scission into •$HO_2$ radicals and conjugated olefins or smaller species which do not lead generally to branching agents. Because of the decrease in the production of peroxides, the reactivity of the alkanes drops off in the 700 to 800 K temperature range. This temperature region is consequently called the Negative Temperature Coefficient (NTC) zone. At higher temperatures, the decomposition of $H_2O_2$:

$$H_2O_2 (+M) = 2OH (+M)$$

opens a new pathway for increased reactivity in the hydrocarbon oxidation system. This reaction becomes very fast above 900 K.

In all the figures presented hereafter, the points refer to experimental observations and the curves to simulations. In order to keep this paper reasonably short, the complete mechanisms mentioned in this paper are not presented here, but are available on request.

EXPERIMENTAL METHOD AND RESULTS

The two series of experiments were conducted using an apparatus, the conception of which is very close to one which has been recently used in our laboratory to study the thermal decomposition of n-dodecane (27).

*Experimental procedure*

Experiments were carried out in a continuous jet stirred reactor (internal volume about 92 cm$^3$) (28) made of quartz and operated at a constant temperature and pressure. The heating of the reactor was achieved by means of electrically insulated resistors directly coiled around the vessel. Temperature was measured by using a thermocouple located inside the reactor in a glass finger. In order to obtain a spatially homogeneous temperature inside the reactor, reactants were preheated at a temperature close to the reaction temperature. Corresponding residence time in the preheating section was approximately 1% of the total residence time.

The liquid fuels (decane and n-hexadecane, both 99% pure) were contained in a glass vessel pressurized with helium. After each load of the vessel, helium bubbling and vacuum pumping were performed in order to remove oxygen traces dissolved in the liquid hydrocarbon fuel. The liquid reactant flow rate was controlled by using a liquid mass flow controller, mixed to the carrier gas (helium, 99.995% pure) and then evaporated by passing through a single pass heat exchanger whose temperature is set above the boiling point of the mixture. Carrier gas and oxygen (99.5% pure) flow rates were controlled by gas mass flow controllers located before the mixing chamber. The fuel-oxygen-helium mixtures were premixed prior the introduction into the reactor.

Due to the formation of compounds, which were either gaseous or liquid at room temperature, the analyses of the products leaving the reactor were performed in three steps:

➢ For the analysis of initial alkanes, the outlet flow was directed towards a trap maintained at liquid nitrogen temperature during a determined period of time. The use of helium instead of nitrogen as dilutant avoids the carrier gas to be trapped and has a negligible effect on the obtained results compared to the use of nitrogen. At the end of this period (typically 16 min), the trap was removed and, after the addition of acetone and of an internal standard (n-octane), progressively heated up. When the temperature of the trap was close of 273 K the mixture was poured into a small bottle and then injected by an auto-sampler in a gas chromatograph with flame ionization detection (FID) for quantification. The column used for the separation was an HP-1 capillary column with helium as carrier gas. Calibrations were performed by analyzing a range of solutions containing known amounts of n-octane and of the studied alkane. Products of the reaction liquid at ambient temperature were not analyzed here.

➢ Oxygen, carbon oxides and $C_1$-$C_2$ hydrocarbons were analyzed on-line by a chromatograph fitted with both thermal conductivity detector (TCD) for oxygen and carbon oxides detection and FID for methane and $C_2$ hydrocarbons detection and using a carbosphere packed column with helium as carrier gas. Species identification and calibration was realized by injection of gaseous standard mixtures provided by Air Liquide. The fact to use helium as carrier gas both in the reactor and in this gas chromatograph facilitates the quantification of oxygen.

➢ For other gaseous products, samples were directly obtained by expansion in an evacuated volume and chromatographically analyzed by FID using a Haysep packed column and nitrogen as carrier gas. The identification of these compounds was performed by comparison of retention times when injecting the product alone in gas phase. Calibrations were performed by analysing a range of samples containing known pressures of each pure compound to quantify. Hydrogen, water and formaldehyde were not quantitatively analysed.

Calculated uncertainties on the species quantifications were about ± 5% with the online analysis of oxygen and $C_1$-$C_2$ hydrocarbons and about ± 10% for the analysis of other species. The carbon balance between the inlet and outlet of the reactor was found to be checked about ± 10% for temperatures above 1000 K, for which the quasi-exclusive formation of light compounds can be assumed.

*Experimental results for the low-temperature oxidation on n-decane*

The experiments were carried out at a mean residence time of 1.5 s, temperatures from 550 to 1100 K, atmospheric pressure and for stoichiometric mixtures containing 0.23% (mol) of n-decane. This low concentration of fuel has been chosen in order to avoid temperature gradients inside the reactor. Figure 2 presents the profiles of reactants and of the main light products, namely carbon oxides, methane, ethylene, ethane, propene, acetaldehyde and $C_3$ aldehydes (propanal and acrolein could not be distinguished with the used chromatographic method). The results exhibit a clear NTC behavior between 600 and 750 K, the start of this NTC zone is shifted around 50 K below compared to the results of Dagaut et al. (13) at 10 bar. This shift was expected due to the influence of pressure on the equilibrium of the addition reactions of molecular oxygen to the alkyl and hydroperoxyalkyl radicals. Below 800 K, the major light products are carbon monoxide, ethylene and propene. In this temperatures range, ethylene and propene are formed in similar amounts, while the formation of the $C_2$ species is predominant at higher temperature. Very small amounts of acetaldehyde and $C_3$ aldehydes are also observed below 800 K. The formation of carbon dioxide, methane and ethane is very low below 850 K, the mole fraction of carbon dioxide being below the detection limit of $2 \times 10^{-4}$ due to the low sensitivity of TCD.

**FIGURE 2**

*Experimental results for the low-temperature oxidation on n-decane/n-hexadecane blend*

These experiments have been performed under the same conditions as for n-decane, with a mixture containing the same number of carbon atoms, with a ratio of 35% (mol) of n-hexadecane in n-decane. The obtained results are displayed in figure 3 and exhibit the same NTC behaviour as in the case of pure n-decane. A comparison between the experimental results obtained in both cases is presented in figure 4 for the extent of conversion and in figure 5 for the formation of light products. At each studied temperature, the extent of conversion of n-decane is similar and close to that of pure n-decane. At temperature above 800 K the extends of conversion of n-hexadecane and n-decane are similar, but the $C_{16}$ compound is slightly more reactive at lower temperatures. The mole fraction of oxygen and the main light products are almost identical for the mixture and for pure n-decane, as illustrated by figures 5a and 5b for carbon monoxide and ethylene. Only the concentrations of methane and ethylene are slightly larger and that of propene (fig 5d) lower in the case of the mixture.

<div align="center">**FIGURES 3-5**</div>

THE GENERATION OF THE MODEL

The EXGAS software has already been extensively described in the case of alkanes (7)(29), as well as for alkenes (30).

*General features of EXGAS*

The system provides reaction mechanisms made of three parts:

➢ A $C_0$-$C_2$ reactions base including all the reactions involving radicals or molecules containing less than three carbon atoms (31),

The kinetic data used in the $C_0$-$C_2$ reactions base were taken from the literature, mainly the values recommended by Baulch et al. (32) and Tsang et al. (33). The pressure-dependent rate constants follow the formalism proposed by Troe (34) and efficiency coefficients have been included.

➢ A comprehensive primary mechanism, where the only molecular reactants considered are the initial organic compounds and oxygen,

The primary mechanism includes only elementary steps; the reactions, which are considered to model the oxidation of alkanes are the following:

♦ Unimolecular and bimolecular initiation steps,

To be in better agreement with values of the literature (35), the rate constants of the unimolecular decompositions involving the breaking of a C-C bond have been multiplied by a factor 3 compared to our previous work (7) and decompositions of alkyl radical by breaking of a C-H bond have been taken into account.

♦ Decomposition and oxidation of alkyl radicals (to form the conjugated alkene and $HO_2^\bullet$ radicals),

All non-substituted alkyl radicals have been considered for oxidation to give alkenes and not only those deriving from the reactant (7).

- Addition to oxygen of alkyl and hydroperoxyalkyl radicals,
- Isomerizations of alkyl and peroxy radicals,
- Decompositions of hydroperoxyalkyl and di-hydroperoxyalkyl radicals to form cyclic ethers, alkenes, aldehydes, ketones and oxohydroperoxyalkanes,
- Metatheses involving the H-abstraction reactions from the initial reactants by a radical,
- Termination steps.

➢ A lumped secondary mechanism, containing reactions consuming the molecular products of the primary mechanism, which do not react in the reaction bases (29).

Thermochemical data for molecules or radicals are automatically calculated and stored as 14 polynomial coefficients, according to the CHEMKIN II formalism (36). These data are calculated using software THERGAS (37), based on the group and bond additivity methods proposed by Benson (38). The kinetic data of the reactions included in the primary or secondary mechanisms are either calculated using thermochemical kinetics methods or estimated using correlations (7)(29).

*Improvements of the generation of reactions of primary products (secondary mechanism)*

As shown in figure 6, reactions of primary products were generated by EXGAS for every type of molecular products (hydroperoxides, alkenes, cyclic ethers, aldehydes, alkanes, ketones, alcohols). In order to have a manageable size, the lumped secondary mechanisms (29) involved lumped reactants: the molecules formed in the primary mechanism, with the same molecular formula and the same functional groups, are lumped into one unique species without distinction between the different isomers. Up to now, these secondary mechanisms included global reactions which mainly produced in the smallest number of steps molecules or radicals, the reactions of which are included in the $C_0$-$C_2$ reactions base. For instance, the addition of •OH radicals to decenes was written whatever the isomer:

$$C_{10}H_{20} + •OH \Rightarrow HCHO + •CH_3 + 4C_2H_4$$

This method of lumping is a simplified version of that proposed by Ranzi et al. (39), which used fractional coefficients. By using this method the addition of •OH radicals to decenes would be written:

$$C_{10}H_{20} + •OH \Rightarrow 0.5 CH_3CHO + 0.5 C_2H_5CHO + 0.8333 •C_7H_{15} + 0.1667 •C_{10}H_{21} \quad (5).$$

The kinetic data of the reactions of primary products generated by EXGAS are those of the first involved reaction as shown in figure 6. A more detailed secondary mechanism had been considered only for cyclic ethers including the same number of carbon atoms as the reactant and including more than 3 atoms in the cycle. In this case, cyclic radicals obtained by H-abstractions were decomposed by beta-scission decomposition and opening of the ring, but could also react with oxygen molecules. The reactions with oxygen involved the classical sequence of oxygen addition, isomerization, second oxygen addition, second isomerization and beta-scission to produce cyclic ether hydroperoxydes, degenerate branching agents, which decomposed to give •OH free radicals and several molecules or radicals, the reactions of which were included in the $C_0$-$C_2$ reactions base.

**FIGURE 6**

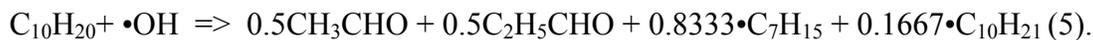

To reduce the large overprediction observed for the formation of ethylene and formaldehyde at low temperature and to better reproduce the reactivity during the low-temperature oxidation of large alkanes at high pressure, an attempt to have a more realistic modeling of the fate of primary products was made. The rules used to generate the reactions of primary products have consequently been modified. The general idea is to promote the formation of $C_{2+}$ alkyl radicals, the reactions of which are already included in the primary mechanism. As these radicals would mainly be obtained by beta-scission decomposition of linear molecules, the formation of alkyl radicals with the radical center carried by the first carbon atom has been considered. In order to more closely model the reactivity of the primary products, an effort was also made for the rate constants of H-abstractions to better reflect the number of different types of abstractable H-atoms present in the considered molecules, i.e. primary and secondary alkylic H-atoms in all compounds, secondary allylic in alkenes (the abstractions of vinylic H-atoms were neglected) and carboxylic H-atoms in aldehydes.

The new rules of generation are given in Tables 2-5. Starting from a decomposition by breaking of a O-OH bond (Table 2), the species produced are carbon monoxide, aldehydes, OH and alkyl radicals from ketohydroperoxides, aldehydes, OH and alkyl radicals from saturated hydroperoxides, acrolein and OH and alkyl radicals from unsaturated hydroperoxides. While alkyl radicals of different sizes can be obtained from the decomposition of hydroperoxides, we consider here only the formation of those containing about half of the atoms of carbon present in the molecules for which the secondary reaction is written. In the case of the H-abstraction/decomposition of cyclic ethers (Table 3) including more than 3 atoms in the cycle, ketones and alkyl radicals are formed. The cyclic ether radicals obtained by H-abstractions from cyclic ethers including the same number of carbon atoms as the reactant and including more than 3 atoms in the cycle can still react with oxygen molecules and ultimately lead to cyclic ether hydroperoxides or be decomposed to give alkene molecules and •$CH_2CHO$ radicals. The decomposition of cyclic ether hydroperoxides leads to alkenes, carbon dioxide, •OH and •HCO radicals. Alkanes, aldehyde, ketones and alcohols (Table 4) react first by H-abstractions and lead to alkyl radicals in the first case, to alkoxy radicals in the second case, to ketene molecules and alkyl radicals in the third case and to alkene and formaldehyde molecules and OH and alkyl radical in the last case. The number of reactions of primary products is larger in the case of alkenes (Table 5): they can react by additions of small radicals, H-abstractions and retro-ene reactions. The addition of H-atoms leads to the formation of alkyl radicals. The addition of •OH radicals produces aldehydes and alkyl radicals. Ketene, H-atoms and alkyl radicals are obtained by addition of O-atoms. Propene or iso-butene and alkyl radicals are formed by addition of •$CH_3$ radicals. Finally, the addition of •$HO_2$ gives epoxides and •OH radicals. The formation of 1,3-butadiene molecules and alkyl radicals are obtained by H-abstractions and propene and that of other alkenes by retro-ene reactions. The values of the rate constants for the additions take into account the possibility of radicals to add on both side of the double bond.

**TABLE 2-5**

COMPARISON BETWEEN EXPERIMENTAL RESULTS AND MODELING

Simulations have been performed using softwares of CHEMKIN II (36): PSR for results obtained in a jet-stirred reactor and SENKIN for data measured in a shock tube and in a flow reactor. To reduce the calculation times, a model specific to the oxidation of n-decane has been generated in addition to that for the mixture.

*Validation using the jet-stirred reactor data obtained in this study*

Simulations in the case of n-decane and the n-decane/n-hexadecane blend are displayed in figures 2 and 3, respectively. These figures show that both models can well reproduce the position of the NTC zone, the amplitude of this phenomenon and the profiles of the concentration of oxygen, carbon oxides, ethylene below 900 K and ethane. However they overestimate the consumption of hydrocarbons below 750 K and above 850 K. The formation of carbon monoxide, methane, ethylene and propene is underestimated above 1000 K. The production of propene below 750 K is well simulated in the case of the mixture, but underestimated for pure n-decane. At 650 K, ethylene and propene are formed by reaction with oxygen molecules of ethyl and propyl radicals radicals, respectively. These radicals derived from the decomposition of hydroperoxides species. In both cases, the production of acetaldehydes is well simulated above 800 K, but considerably overestimated below 750 K. The formation of $C_3$ aldehydes is overestimated both below 750 K and above 800 K. The simulation indicates that below the NTC zone, propanal, which is obtained by decomposition of ketohydroperoxides, is the major $C_3$ aldehyde, while it is acrolein at higher temperature. Acrolein is produced from $C_3H_5OOH$ which is formed by combination of allyl and $HO_2$• radicals. Figures 4 and 5 present a comparison between the simulations obtained for pure n-decane and for the mixture. The similarity in the extent of conversion of n-decane in both cases and in that of both hydrocarbons present in the mixture above 800 K is correctly modeled, as well as the slightly higher reactivity of n-hexadecane at lower temperatures. Simulations reproduce also well the similarity in the formation of light products experimentally observed: the curves can almost not be distinguished for the production of carbon monoxide and propene. Only a small difference is obtained for that of ethylene (at 700 K, its formation is 19% lower in the case of the mixture) and acetaldehyde.

The fact that the rate of fuel consumption is overestimated below 750 K while many other light intermediates are satisfactorily predicted indicates that the model should also overestimate the formation of heavier products (e.g. cyclic ethers), the analysis of which would be of great interest in a future work. This overestimation of the low-temperature reactivity could be due to the fact that we consider the immediate decomposition to give aldehydes and alkyl radicals of the alkoxy radicals obtained by decomposition of hydroperoxides.

*Validation using the jet-stirred reactor data obtained by Dagaut et al. (13)*

Dagaut et al. (13) have measured the profiles of a wide range of species during the oxidation of n-decane in a jet-stirred reactor for temperatures from 550 to 1050 K and pressures equal to 10 bar. Experiments and simulations are presented in figure 7. As for the data obtained at 1 bar, the model can well reproduce the position of the NTC zone, but the amplitude of the phenomenon is much underestimated in the case of the profile of n-decane, the concentration of which is substantially underestimated below 800 K. The prediction of the profiles of the concentration of oxygen, carbon dioxide, carbon monoxide below 900 K, ethylene, ethane (not shown here), formaldehyde and acetaldehyde is satisfactory. The formation of larger alkenes and aldehydes is underestimated below 850 K. However, the repartition between the major large alkenes is well reproduced. These unsaturated compounds are mainly formed by beta-scission from large alkyl radicals.

**FIGURE 7**

Figure 7 displays also results of a simulation made with the previous mechanism of Buda et al. (7), i.e. before the improvements proposed in the present work, for the compounds for which the largest effect has been observed. With this older mechanism, the amplitude of the NTC phenomenon was still underpredicted, but the concentration of n-decane was much less underestimated. On the contrary, the prediction of the profiles of oxygen, carbon monoxide, ethylene, acetaldehyde and propene were much less satisfactory. At 650 K, the concentration of ethylene was overpredicted by a factor 10. This positive effect is completely due to the new rules proposed for writing reactions of primary products. As illustrated in the case of 1-heptene, the effect on large alkenes is balanced: in the two simulations the predicted formation is too low below 750 K, the profile obtained with the mechanism of Buda et al. is better between 750 and 850 and the profile computed with the new mechanism is closer to the experimental results at higher temperature. The lack of $C_{2+}$ alkenes in the lower temperatures range shows that, while the new rules proposed for writing reactions of primary products has led to a considerable improvement of the prediction of the formation of products, these simplified rules are not able to completely reproduce the complexity of the low-temperature chemistry. In the higher temperatures range, the fact to consider in more details the different H-atoms which can be abstracted (primary or secondary, allylic and alkylic) allows the consumption of large alkenes to be better reproduced. At 850 K, the abstractions of alkylic H-atoms, which were not considered previously account for 60 % of the total H-abstractions from 1-heptene. The importance of retro-ene reactions starts only to be noticeable above 1000 K. At 1050 K, 8 % of 1-heptene is consumed through this way.

*Validation using shock tube data*

Pfahl et al. (11) measured ignition delay times for n-decane/oxygen/nitrogen mixtures in a shock tube for temperatures from 660 to 1200 K and pressures equal to 12 and 50 bar. Zhukov et al. (14) have performed similar experiments but for pressures equal to 13 and 80 bar. Experiments and simulations are presented in figure 8. First it should be noted that at 12-13 bar, the delays measured by Zhukov et al. (14) are shorter by a factor about 2 than those obtained by Pfahl et al. (11). That explains why at this pressure our model is in good agreement with the results of Pfahl et al. (11), but oversestimates those of Zhukov et al. (14). Our improved model can more satisfactorily reproduce the results at 50 and 80 bar. Between 850 and 1000 K, the results computed by the new mechanism are shorter by a factor about 2 than the values obtained with the mechanism of Buda et al. (7).

**FIGURE 8**

*Validation using turbulent flow tube data*

Lenhert et al. (19) have studied the oxidation of n-dodecane in a flow reactor pressurized at 8 bar. They have followed the fuel consumption and the formation of few intermediates using the control cool down method (CCD), in which the reactor was stabilized to a specified maximum temperature and then allowed to cool at a fixed rate and a constant pressure and residence time. Experiments and simulations are displayed in figure 9 for the mole fractions of n-dodecane and the major products. The consumption of fuel is underestimated below 680 K and overestimated above. The computed NTC behavior above 650 K is much less pronounced than that experimentally observed. The prediction of the formation of carbon dioxide is relatively good, while the profiles of carbon monoxide and ethylene are reproduced within a factor about 2. The prediction of the amounts of produced methane, propene and 1-butene is satisfactory, but an inflexion point in the mole fraction profiles is shown by the simulations and not experimentally observed. The simulations made by the authors using the mechanism of the team of Ranzi and Faravelli (18) led also to rather passable agreements.

**FIGURE 9**

DISUCUSSION ABOUT THE INFLUENCE OF REACTIONS OF PRIMARY PRODUCTS

Figures 10 and 11 presents flow rates and sensitivity analyses, respectively, performed at 650 and 900 K for the oxidation of n-decane at atmospheric pressure under jet-stirred reactor conditions. Figure 12 presents a sensitivity analysis for the autoignition delay times of this reactant under shock tube conditions at 12 bar. Since a sensitivity analysis of the effects of primary reactions on the autoignition of n-heptane was already presented in a previous paper (7), the objective of the present sensitivity analysis is only to better understand the influence of the considered reactions of primary products. Sensitivity analyses have been obtained by multiplying by a factor 10 the rate constant of each generic reaction.

**FIGURES 10-12**

At 650 K, the main primary products are cyclic ethers, ketones and ketohydroperoxides, and to a lesser extent alkenes and saturated hydroperoxides, all these compounds including 10 atoms of carbons. At this temperature, the decomposition of hydroperoxides is the only important type of reactions as regards ignition delay times, while a variation of its rate constant has a negligible effect on species mole fractions in a jet-stirred reactor. As it consumes the most abundant primary products, the reactions of primary products of cyclic ethers favors the formation of smaller compounds, such as carbon oxides, ethylene, propene and $C_{1-2}$ aldehydes. The decomposition of cyclic ethers radicals competes with their addition to oxygen molecules, which yields cyclic ether ketohydroperoxides. An increase of the rate constant of this class of reaction counteracts then the increase of the extent of n-decane conversion and of the formation of carbon dioxide induced by the decomposition of these cyclic ketohydroperoxides. The formation of propanal in influenced by the decomposition of ketones, because the H-abstractions from decanone leads to the formation of octyl radicals and the decomposition of $C_8$ ketohydroperoxides produces this $C_3$ aldehyde. Large aldehydes being produced by the decomposition of ketohydroperoxides and the addition of •OH radicals to alkenes, their reactions have an effect on the formation of smaller compounds, such as carbon monoxide, ethylene, propene and lighter aldehydes. The H-abstractions from alkenes lead to the formation of 1,3-butadiene, which react mainly by addition of •OH radicals to produce formaldehyde and allyl radicals. The formation of these resonance stabilized radicals which are the main source of acrolein explains the retarding effect of the H-abstractions from alkenes on the reactivity and the promoting one on the production of acrolein. Propanal is partly obtained by addition of •OH radicals to 1-butene, which explains that its production is enhanced when the rate constants of the additions to alkenes are increased. H-abstractions and additions being the two competing consumption channels of alkenes, the sensitivity analysis displays opposite effects for each of them. At this low temperature, retroene reactions are negligible.

At 900 K, 95% of the primary products are alkenes containing less than 10 atoms of carbon. $C_{10}$ alkenes, ketones and cyclic ethers are minor products, as well as unsaturated hydroperoxides containing less than 10 atoms of carbon and obtained by β-scission decomposition from hydroperoxyalkyl radicals. At this temperature, only the reactions of primary products of alkenes and aldehydes, which are obtained from alkenes by addition of •OH radicals, have a noticeable effect on ignition delay times. In a jet-stirred reactor, only changes in the rate constants of the reactions of alkenes lead to important variations in species mole fractions. Like at 650 K, the formation of allyl radicals, which derives from the H-abstractions from alkenes, explains that an increase of the rate constant of this reaction leads to a smaller reactivity and formation of products, but favors the formation of acrolein. The competing channel, the additions of small radicals to the double bond, has the opposite effect. Retroene reactions favors the formation of propene, which is obtained by each of them, and to a lesser extent that of acetaldehyde and propanal, which are obtained addition of •OH radicals to propene and 1-butene.

At 1100 K, the influence of reactions of primary products on autoignition delay times is lower as the reactivity is mainly driven by reactions of the $C_0$-$C_2$ reactions base, such as the reactions of $HO_2$• radicals. The uncertainties on some of these reactions partly explain the disagreements observed when modeling shock tube experiments above 900 K, as shown in fig. 8. At this temperature, an effect of retroene reactions on ignition delay times start only to be noticeable.

The sensitivity analyses presented in this section show of the influence of the rate constant of the reactions of primary products, but not of the way these reactions are written in terms of products. It is worth noting the important influence of the lumping of primary products and of their reactions, as different isomers of a given primary product may yield different types of species. For instance, while a variation of the rate constant of the decomposition of hydroperoxides has a negligible effect on species mole fractions in a jet-stirred reactor, the change in the writing of these reactions which is presented in this paper has a considerable effect on the formation of some products, such as ethylene, as shown in figure 7. This analysis shows that a better knowledge of the chemistry involved during the low-temperature oxidation of oxygenated compounds, such as cyclic ethers or large ketones or aldehydes, is needed to better predict the formation of the products formed during the oxidation of alkanes.

COMPARISON BETWEEN THE REACTIVITY OF LINEAR ALKANES FROM $C_8$ to $C_{16}$

While gasoline and diesel fuel have a "near-continuous spectrum" of hydrocarbon constituents, surrogates composed of a limited number of components have to be defined in order to develop detailed kinetic models. This need of a surrogate is made even more important by the fact that the numbers of species and reactions increase drastically with the number of carbon atoms in the reacting molecule as shown in figure 13. It shows that when going from n-decane to n-hexadecane the numbers of species and reactions are multiplied by a factor about 3. The fact that the same rules of generation are used for each compound makes that there is almost a linear variation of the numbers of species and reactions with the number of carbon atoms in the reacting molecule.

# FIGURE 13

In order to help modelers to define the best surrogate for large alkanes, it is of interest to compare the reactivity of a wide range of them both in terms of the extent of conversion and the formation of light products in a jet-stirred reactor, and for ignition delay times, as it has already been made by Westbrook et al. (20).

*Comparison of the reactivity of large alkanes in a jet-stirred reactor*

The oxidation of several linear alkanes from $C_{10}$ to $C_{16}$ has been simulated under the same conditions as in figure 2, i.e. under jet-stirred reactor conditions at atmospheric pressure, with the initial mole fraction of carbon atoms kept constant (i.e. the initial mole fraction of n-decane is 1.6 times larger than that of n-hexadecane). Figure 14 presents the obtained variation of the extent of conversion and mole fractions of carbon monoxide and ethylene with the number of carbon atoms in the reacting molecules. It shows that, at 700 K, the extent of conversion increases significantly with the size of the reactant (with a smaller increase above $C_{12}$), while at 900 K it does not vary much. This result is in agreement with our experimental observations in the case of the n-decane/n-hexadecane blend. It has been shown previously (7) that, at 700 K in the NTC zone, the isomerizations of peroxy radicals have a great influence on the reactivity. As the rate constants of isomerization increases with the size of the involved transition state (the considered sizes of the ring are from 4 to 8 members) (7), this type of reaction would be favoured when the size of the reactant increases. Another reason could also consist in an effect of the reactions with $O_2$ of the alkyl radicals formed by reactions of primary products: for alkyl radicals larger than $C_3$, these reactions easily lead to the formation of hydroperoxides and have a promoting effect. When the size of the reactant increases, more successive formations of these radicals are then obtained with an accelerating effect on the reactivity. At 900 K, as shown by Buda et al. (7) in the case of n-heptane, the most important reactions are metatheses. For larger alkanes, the rate constants of these reactions can be considered in a first approximation proportional to the number of H-atoms and consequently to that of carbon atoms. At constant mole fraction of carbon atoms, the reactivity should therefore, as observed here, remain constant when increasing the size of the reactant. While the variation of the mole fraction of carbon monoxide at both temperatures and that of ethylene at 900K is small, the mole fraction of ethylene exhibits a sharp decrease (a factor 1.7) between n-decane and n-hexadecane. At 700 K, the decrease of the ethylene mole fraction while the extent of conversion increases is mainly due to the fact that, according to the rules of writing reactions of primary products above described, the alkyl radicals obtained by decomposition of the ketohydroperoxides (see Table 2) directly deriving from n-decane are 1-pentyl radicals. An important source of ethylene is the decomposition of the derived ketohydroperoxides followed by the oxidation of the obtained ethyl radicals.

# FIGURE 14

*Comparison of the reactivity of large alkanes for ignition delay times*

The ignition of several linear alkanes from $C_7$ to $C_{16}$ has been simulated under the same conditions as in figure 8, i.e. under shock tube conditions and stoichiometric alkane/air mixtures at 12 bar. The initial hydrocarbon mole fractions are very close to that obtained with the initial mole fraction of carbon atoms kept constant. The obtained evolution of ignition delay times with temperature are displayed in figure 15. As observed by Westbrook et al. (20), the values of delay times of all studied compounds are very close for temperatures below 750 and above 1000 K with more marked difference in between. The results of the team of Livermore indicate that below 1000 K ignition times increase when the size of the reactant increases, while it is the contrary above this temperature, and that the position of the NTC zone does not vary with the studied compound. In our case, when the size of the reactant increases, delay times decrease whatever the temperature and the position of the end of the NTC zone is shifted towards higher temperatures. In the Livermore study, at 800 K, a much larger difference is obtained between n-hexadecane and n-decane than between n-decane and n-octane. Figure 15 exhibits a really opposite trend with only small differences computed between alkanes from $C_{10}$ to $C_{16}$. The experimental results measured by the team of Adomeit for n-heptane (42) and n-decane (11) show a close reactivity between the two fuels but with slightly lower ignition delay times for n-decane in all the temperature range with a more marked difference between 800 and 900 K. An increase of the reactivity with the size of the initial reactant is in agreement with the results presented in the previous paragraph for jet-stirred reactor conditions.

**FIGURE 15**

An artifact due to the way secondary mechanisms are generated is worth noting. Figure 16 presents the variation of simulated cool flame and ignition delay times vs. the number of carbon atoms in the reactant for a series of linear alkanes from $C_7$ to $C_{16}$. It shows well the decrease of these delay times when the size of the linear chain increases, but while an almost monotonic behavior is obtained for cool flame and ignition at 750 K, a much more broken line is obtained for ignition delay times at 900 K. At this temperature, the reactions of the compounds formed in the secondary mechanism (e.g. aldehydes as shown in fig. 12) have a more important influence than at lower temperature and the difference in the writing of the reactions of primary products between compounds including an odd or an even number of carbon atoms explains the observed discontinuities. This large sensitivity of the reactions of the compounds formed in the secondary mechanism, which is shown at 900 K, the temperature for which the increase of the reactivity with the size of the reactant is the largest, would support the fact that this simulated increase is related with the way reactions of primary products are written. New experimental results would be of interest to know the actual variation of the reactivity with the size of the reactant.

**FIGURE 16**

As shown by figure 17, the presence of n-hexadecane in a n-decane/n-hexadecane blend has a promoting effect on ignition delay times at 900 K, the temperature for which the difference observed between both fuels is the largest. For a given concentration, the ignition delay time of the mixture is lower than the average of the delay times of each constituent. Therefore simulating the behavior of a mixture of large alkanes ($\geq C_{10}$) by using n-decane as model compound would lead to an underprediction of the reactivity.

**FIGURE 17**

CONCLUSION

New experimental results have been measured for the oxidation of n-decane and a 65 % (mol) n-decane/ 35 % n-hexadecane blend in a jet-stirred reactor at atmospheric pressure for temperatures from 550 to 1050 K, including the NTC zone. These results show a slightly larger reactivity of the $C_{16}$ compound compared to n-decane, but very similar results for the formation of light products in both studies. Additional experiments including the quantification of heavier products and allowing a carbon balance to be checked would be of great interest and are planned to be performed by our group in a near future.

New detailed kinetic mechanisms for the oxidation of large alkanes at low and intermediate temperature have been proposed. They have been obtained using an improved version of software EXGAS in which new rules of generation of the reactions of primary products have been implemented. These mechanisms lead to a good simulation of the experimental results presented in this paper, as well as of data from the literature obtained in a jet-stirred reactor at 10 atm, in shock tubes for pressures between 12 and 80 bar and in a flow reactor at 8 atm.

Flow rates and sensitivity analyses performed in order to better understand the influence of reactions of primary products have shown that a better knowledge of the chemistry involved during the low-temperature oxidation of oxygenated compounds, such as cyclic ethers or large ketones or aldehydes, would be particularly valuable to better predict the formation of the products formed during the oxidation of alkanes.

Mechanisms have also been generated for a range of n-alkanes from $C_7$ to $C_{16}$ and used to simulate two test cases: the oxidation reactivity under jet-stirred reactor conditions at atmospheric pressure and auto-ignition delay times under shock tube conditions at 12 bar. In each case, the initial mole fraction of atom of carbon was kept constant. In the first case, close extent of conversion and formation of carbon monoxide were obtained whatever the alkane between $C_{10}$ and $C_{16}$. In the second case, ignition delay times increased when the size of the reactant increased, but there was less than a factor two difference between n-decane and n-hexadecane whatever the temperature. According to the application sought, these simulations should help users of kinetic mechanisms to know if n-decane is a good enough surrogate to represent diesel fuel or if it is necessary to consider heavier model molecules knowing that the number of included species and reactions will then be considerably increased. However due to the uncertainties on the used mechanisms, additional experimental results are needed to support these simulations.

ACKNOWLEDGMENTS

This work has been supported by ADEME-PREDIT in collaboration with IFP, TOTAL and PSA Peugeot Citroën through the program BIOKIN. The authors wish to thank Dr. N.P. Cernansky for providing full data of their experimental results on n-dodecane and Dr. C.K. Westbrook from Livermore for providing their paper before its submission to Combustion&Flame.

**TABLE 1: Summary of the main experimental results concerning the auto-ignition and the oxidation of linear alkanes from $C_{10}$ below 900 K, as well as of the models used to simulate them.**

| Compounds | Type of reactor | Temperature range (K) | Pressure range (bar) | Equivalence ratio range | References | References of the related models |
|---|---|---|---|---|---|---|
| n-decane | Shock tube | 650-1200 | 12-50 | 1-2 | Pfahl et al. (1996) (11) | Chevalier et al. (1996)(10), Buda et al. (2005)(7), Ranzi et al. (2005)(15), Westbrook et al. (2007)(20) |
| | | 800-1100 | 12-80 | 0.5-1 | Zhukov et al. (2005) | Buda et al. (2005)(7), Westbrook et al. (2007)(20) |
| | Rapid compression machine | 630-706 | 7-30 | 0.8 | Kumar et al. (2007) (21) | Westbrook et al. (2007)(20) |
| | Jet-stirred reactor | 550-1200 | 10 | 0.3-1.5 | Dagaut et al. (1994, 95, 2002) (13)(17)(22) | Battin-Leclerc et al. (2000)(12), Ranzi et al. (2005)(15), Westbrook et al. (2007)(20) |
| n-dodecane | Flow reactor | 600-800 | 8 | 0.2-0.3 | Cernansky et al. (2004, 05) (18)(19) | Ranzi et al. (2005)(15), Westbrook et al. (2007)(20) |

**TABLE 2: New rules for the generation by EXGAS of the reactions of primary products of acyclic hydroperoxides. Rate constants are given in $cm^3$, $s^{-1}$, mol units.**

| Type of reaction | Obtained products | Rate constant |
|---|---|---|
| *Ketohydroperoxides ($C_nH_{2n}O_3$)* | | |
| Decomposition by breaking of a O-OH bond | n is even (n ≥ 4): <br> •OH+CO+•$C_{n/2}H_{(n+1)}$+ $C_{(n-4)/2}H_{(n-3)}$CHO <br> n is odd (n ≥ 3): <br> •OH+CO+•$C_{(n-1)/2}H_n$+$C_{(n-3)/2}H_{(n-2)}$CHO | $1.5 \times 10^{16}$ exp(-21600/T) <br> (7) |
| *Saturated hydroperoxides ($C_nH_{(2n+2)}O_2$)* | | |
| Decomposition by breaking of a O-OH bond | n is even (n ≥ 2): <br> •OH+ •$C_{n/2}H_{(n+1)}$ + $C_{(n-2)/2}H_{(n-1)}$CHO <br> n is odd (n ≥ 3): <br> •OH+ •$C_{(n+1)/2}H_{(n+2)}$ + $C_{(n-3)/2}H_{(n-2)}$CHO | $1.5 \times 10^{16}$ exp(-21600/T) <br> (7) |
| *Unsaturated hydroperoxides ($C_nH_{2n}O_2$)* | | |
| Decomposition by breaking of a O-OH bond | •OH+ •$C_{n-3}H_{(2n-5)}$ + $C_2H_3$CHO | $1.5 \times 10^{16}$ exp(-21600/T) <br> (7) |

**TABLE 3: New rules for the generation by EXGAS of the reactions of primary products of cyclic ethers and derived species. Rate constants are given in $cm^3$, $s^{-1}$, mol units. Small radicals involved in H-abstractions are •H, •OH, •HO$_2$, •CH$_3$, •CH$_3$OO, •C$_2$H$_5$**

| Type of reaction | Obtained products | Rate constant |
|---|---|---|
| *Cyclic ethers ($C_nH_{2n}O\#x$, x is the size of the ring)* | | |
| H-abstraction by small radicals | n is even (n ≥ 4):  •$C_{n/2}H_{(n+1)}$ + $C_{(n-2)/2}H_{(n-2)}$CO<br>n is odd (n ≥ 5): •$C_{(n+1)/2}H_{(n+2)}$ + $C_{(n-3)/2}H_{(n-3)}$CO<br>x ≥ 4 when deriving from reactant<br>$C_nH_{(2n-1)}O\#x$ | As for the abstraction 6 secondary H-atoms from initial alkane (7) |
| *Cyclic ethers radicals ($C_nH_{(2n-1)}O\#x$)* | | |
| Beta-scission | n ≥ 4:  •CH$_2$CHO + $C_{(n-2)}H_{(2n-4)}$ | $5.0 \times 10^{13}$ exp(-12480/T) |
| *Cyclic ether ketohydroperoxides ($C_nH_{2n-2}O_4$)* | | |
| Decomposition by breaking of a O-OH bond | n ≥ 3:  •OH + CO$_2$ + •HCO + $C_{(n-2)}H_{(2n-4)}$ | $1.5 \times 10^{16}$ exp(-21600/T)<br>(7) |

**TABLE 4: New rules for the generation by EXGAS of the reactions of primary products of ketones, alkanes, aldehydes and alcohols. Rate constants are given in $cm^3$, $s^{-1}$, mol units. Small radicals involved in H-abstractions are •H, •OH, •HO$_2$, •CH$_3$, •CH$_3$OO, •C$_2$H$_5$**

| Type of reaction | Obtained products | Rate constant |
|---|---|---|
| *Alkanes ($C_nH_{2n+2}$)* | | |
| H-abstraction by small radicals | •C$_n$H$_{(2n+1)}$ | As for the abstraction of 6 primary and (2n-4) secondary H-atoms from initial alkane (7) |
| *Aldehydes ($C_nH_{(2n+1)}CHO$)* | | |
| H-abstraction by small radicals | •C$_{(n+1)}$H$_{(2n+1)}$O | As for the abstraction of an aldehydic H-atom[a] |
| *Ketones ($C_nH_{2n+2}CO$)* | | |
| H-abstraction by small radicals | CH$_2$CO + •C$_{(n-1)}$H$_{(2n-1)}$ | As for the abstraction of 6 primary and (2n-4) secondary alkylic H-atoms from initial alkane (7) |
| *Keto radicals ($C_nH_{(2n+1)}CO$)* | | |
| Beta-scission | CO + •C$_n$H$_{(2n+1)}$ | $1.8 \times 10^{14} \exp(-7850/T)$ |
| *Alcohols ($C_nH_{(2n+1)}OH$)* | | |
| H-abstraction by small radicals | •OH + C$_n$H$_{2n}$ | As for the abstraction of 3 primary and (2n-2) secondary alkylic H-atom from initial alkane (7) |
| H-abstraction by small radicals | HCHO + •C$_{(n-1)}$H$_{(2n-1)}$ | As for the abstraction of 1 tertiary alkylic H-atom from initial alkane (7) |

[a]: H : k = 4.0.10$^{13}$ exp(-2100/T); OH : k = 4.2.10$^{12}$ exp(-250/T); HO$_2$ : k = 1.0.1012 exp(-5000/T); CH$_3$ : k = 2.0.10$^{-6}$ exp(-1250/T); C$_2$H$_5$ : k= 1.3.10$^{12}$ exp(-4300/T).

**TABLE 5: New rules for the generation by EXGAS of the reactions of primary products of alkenes ($C_nH_{2n}$). Rate constants are given in $cm^3$, $s^{-1}$, mol units.**

| Other species involved | Obtained products | Rate constant |
|---|---|---|
| *Additions* | | |
| H-atoms | $\bullet C_nH_{(2n+1)}$ | $1.32.10^{13} \exp(-785/T)$ (40) <br> + <br> $1.32.10^{13} \exp(-1640/T)$ (40) |
| OH radicals | $HCHO + \bullet C_{(n-1)}H_{(2n-1)}$ <br> $CH_3 + C_{(n-2)}H_{(2n-3)}CHO$ | $1.4 \times 10^{12} \exp(520/T)$ (40) <br> $1.4 \times 10^{12} \exp(520/T)$ (40) |
| O-atoms | $CH_2CO + H + \bullet C_{(n-2)}H_{(2n-3)}$ | $6.0 \times 10^4 T^{2.56} \exp(770/T)$ (40) |
| $CH_3$ radicals | $C_4H_8 + \bullet C_{(n-3)}H_{(2n-5)}$     $C_3H_6$ <br> $+ \bullet C_{(n-2)}H_{(2n-3)}$ | $1.7 \times 10^{11} \exp(3720/T)$ (40) <br> $9.6 \times 10^{10} \exp(4030/T)$ (40) |
| $HO_2$ radicals | $\bullet OH + C_nH_{2n}O\#3$ | $1.0 \times 10^{12} \exp(7150/T)$ (40) |
| *H-abstractions* | | |
| Small radicals | $n \geq 4: C_4H_6 + \bullet C_{(n-4)}H_{(2n-7)}$ | As for the abstraction of 2 secondary allylic(40), 3 primary alkylic (7) and (2n-8) secondary alkylic (7) H-atoms from initial alkene |
| *Retro-ene decompositions* | | |
| - | $n \geq 5: C_3H_6 + C_{(n-3)}H_{(2n-6)}$ | $8.0 \times 10^{12} \exp(-28400/T)$ (41) |

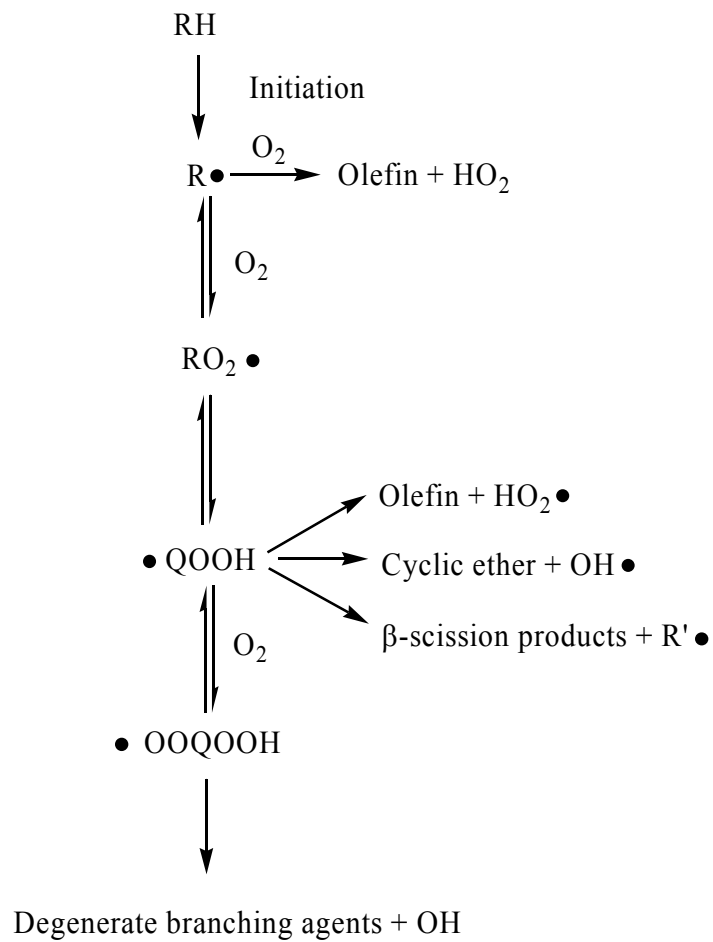

Figure 1: General scheme of the low-temperature oxidation of an alkane (RH).

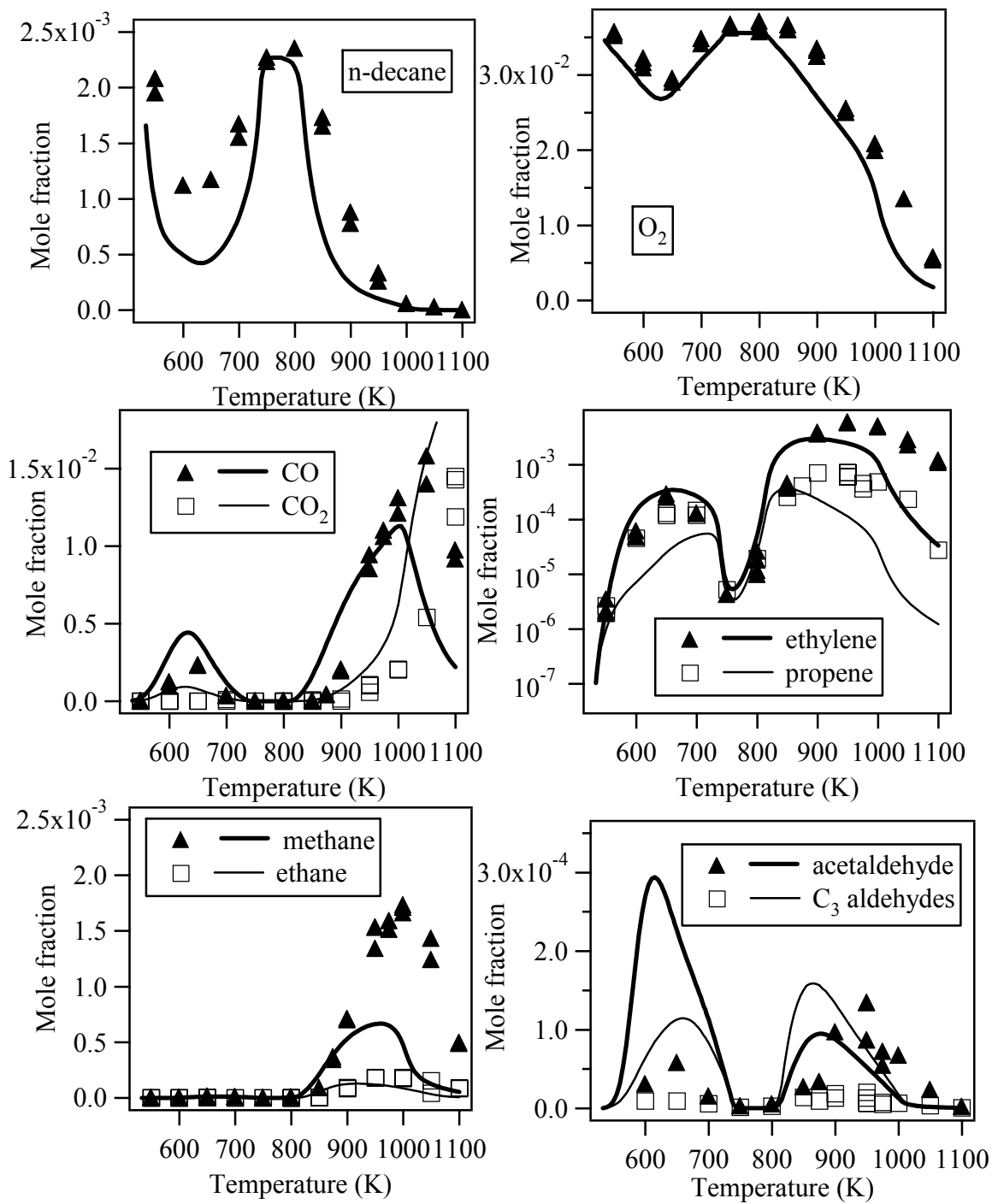

Figure 2: Experimental and computed profiles of species for the oxidation of n-decane in a jet-stirred reactor (residence time of 1.5 s, atmospheric pressure, stoichiometric mixtures containing 0.23% (mol) n-decane diluted in helium).

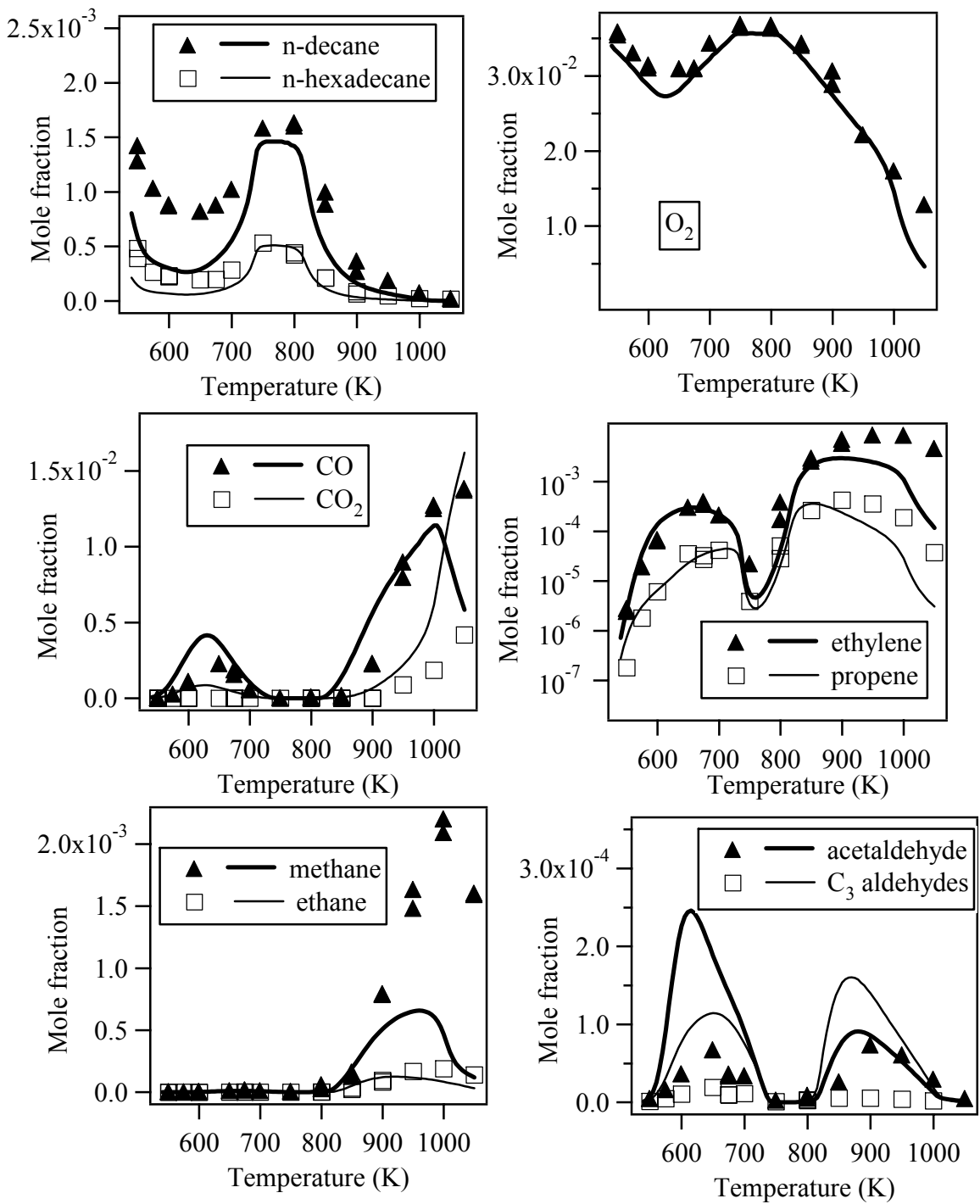

Figure 3: Experimental and computed profiles of species for the oxidation of a n-decane/n-hexadecane blend) in a jet-stirred reactor (residence time of 1.5 s, atmospheric pressure, stoichiometric mixtures containing 0.148 % (mol) n-decane and 0.052% (mol) n-hexadecane diluted in helium).

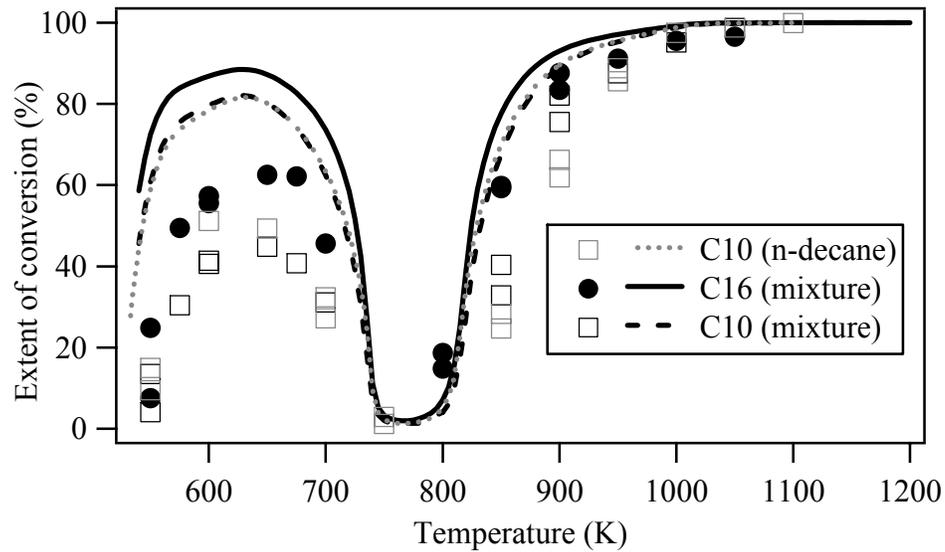

Figure 4: Comparison between the experimental and computed profiles of extent of conversion during the oxidation of pure n-decane and of a n-decane/n-hexadecane blend in a jet-stirred reactor (residence time of 1.5 s, atmospheric pressure, stoichiometric mixtures containing each the same concentration of carbon atoms).

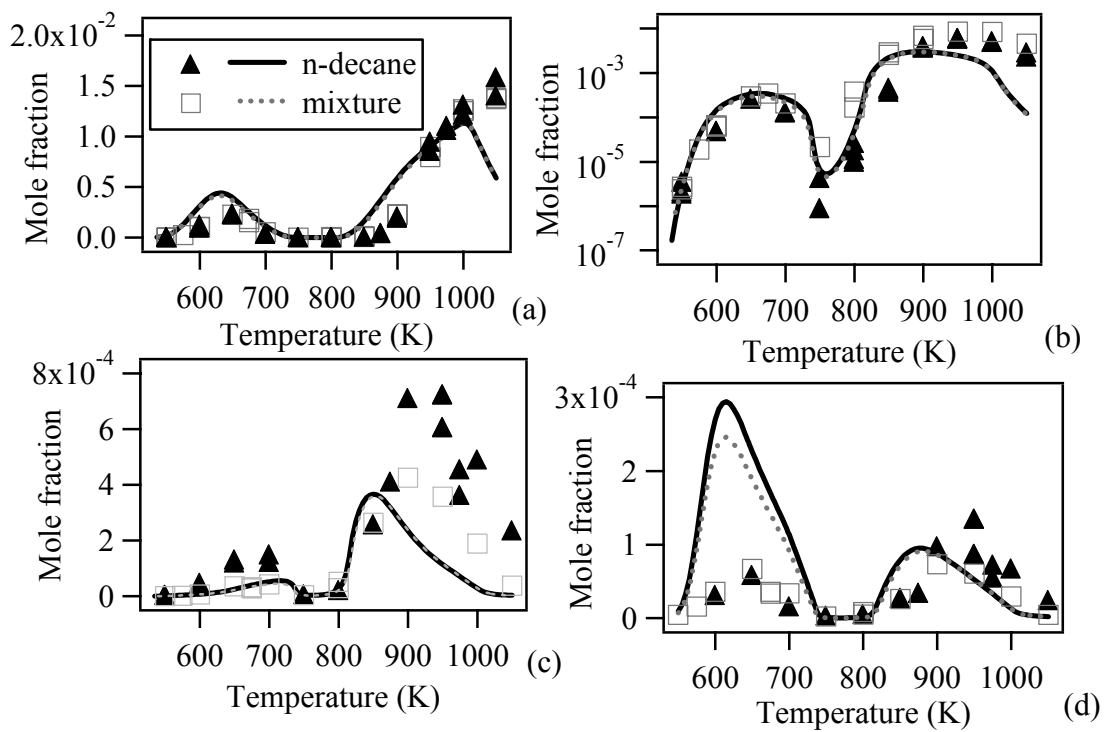

Figure 5: Comparison between the experimental and computed profiles of (a) carbon monoxide, (b) ethylene, (c) propene and (d) acetaldehyde during the oxidation of pure n-decane and of a n-decane/n-hexadecane blend in a jet-stirred reactor (residence time of 1.5 s, atmospheric pressure, stoichiometric mixtures containing each the same concentration of carbon atoms).

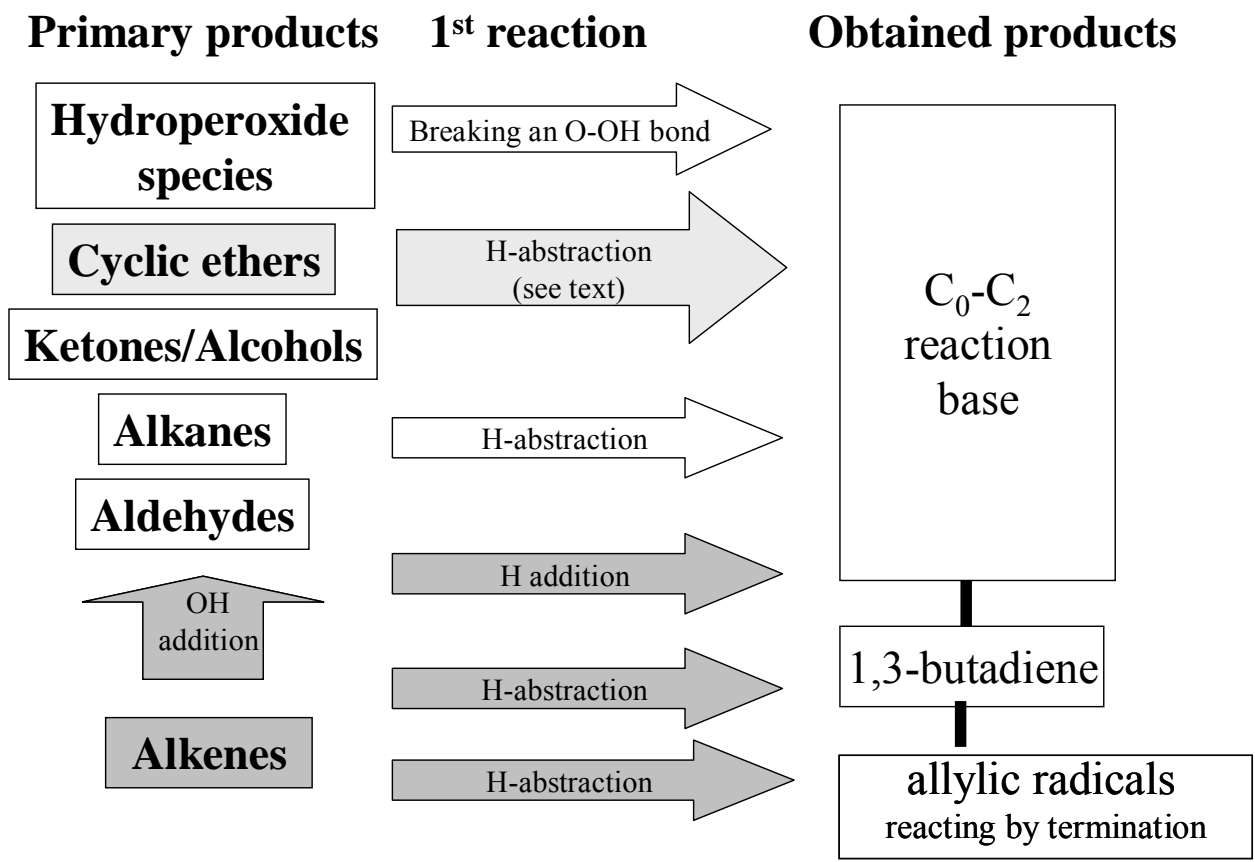

Figure 6: General structure of the previous secondary mechanisms generated by EXGAS.

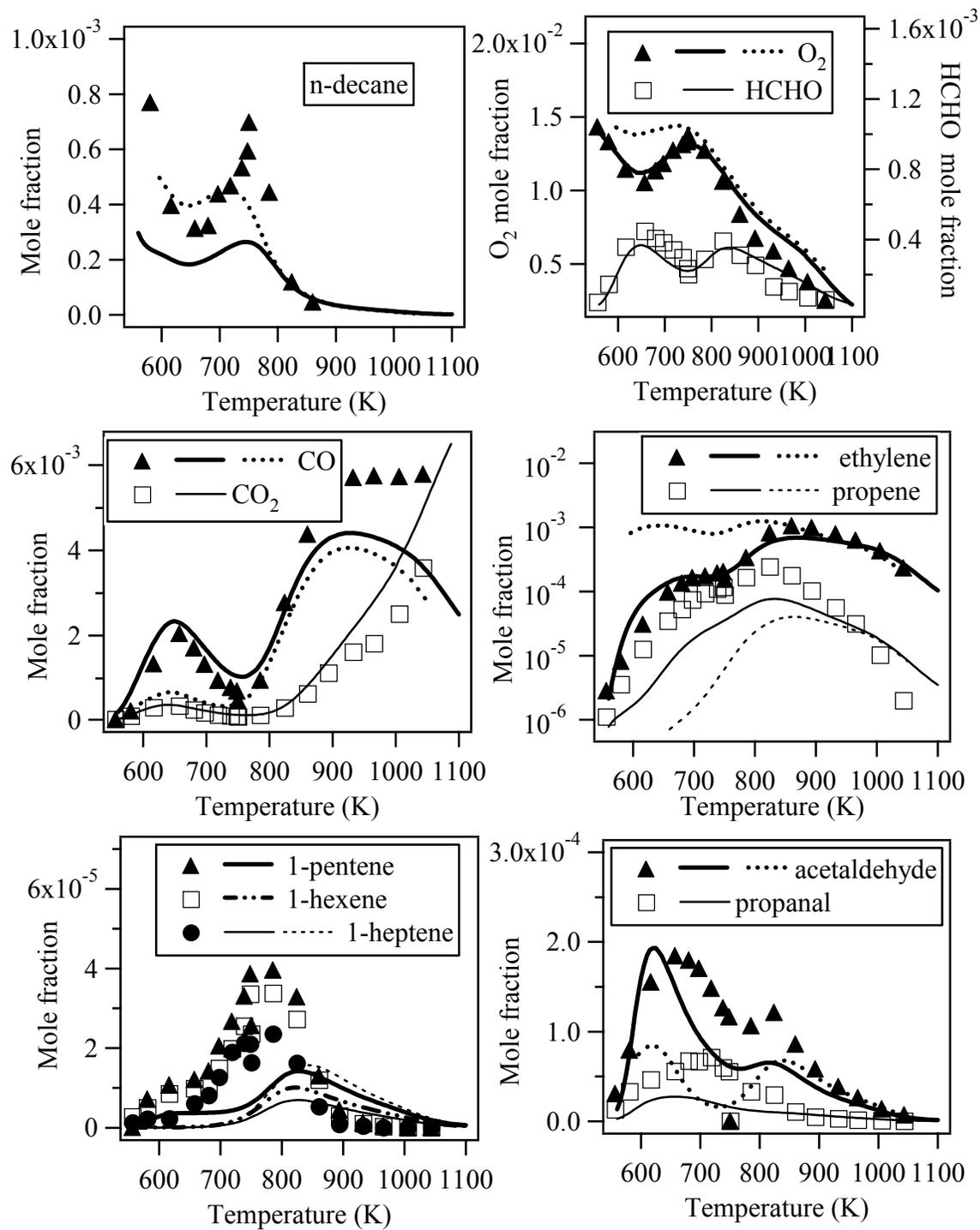

Figure 7: Comparison between computed profiles of species (full lines) and the experimental ones of Dagaut et al. (13) for the oxidation of n-decane in a jet-stirred reactor (residence time of 1 s, 10 atm, stoichiometric initial mixtures containing 0.1% (mol) n-decane). Broken lines correspond to a simulation using the previous mechanism of Buda et al. (7).

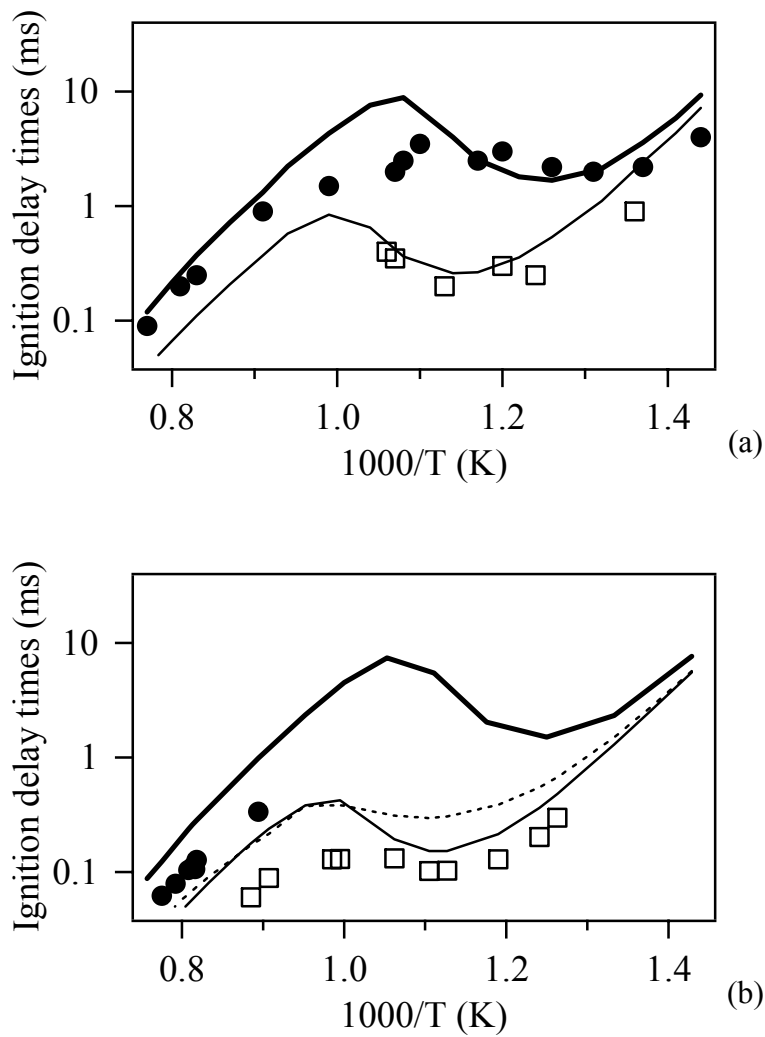

Figure 8: Comparison between computed (full lines) and experimental ignition delay times obtained in shock tubes for stoichiometric n-decane/air mixtures by (a) Phahl et al. (11) at 12 (black circles and thick line) and 50 (white squares and thin line) bar and by (b) Zhukov et al. (14) at 13 bar (black circles and thick line) and 80 bar (white squares and thin line, the thin broken line corresponding to a simulation using the previous mechanism of Buda et al. (7)).

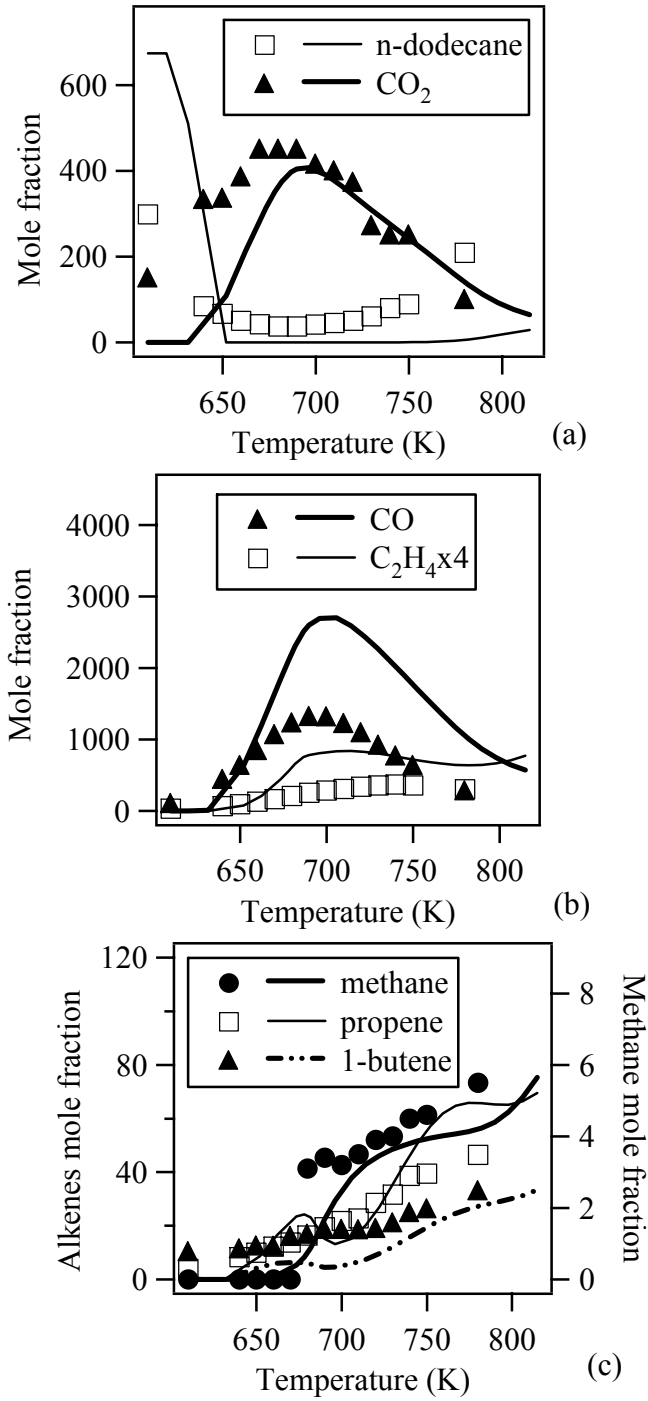

Figure 9: Comparison between computed profiles of species and the experimental ones of Lenhert et al. (19) (in ppm) for the oxidation of n-dodecane in a turbulent (residence time of 125 ms, 8 atm, for initial mixtures containing 674 ppm n-dodecane for $\phi = 0.25$).

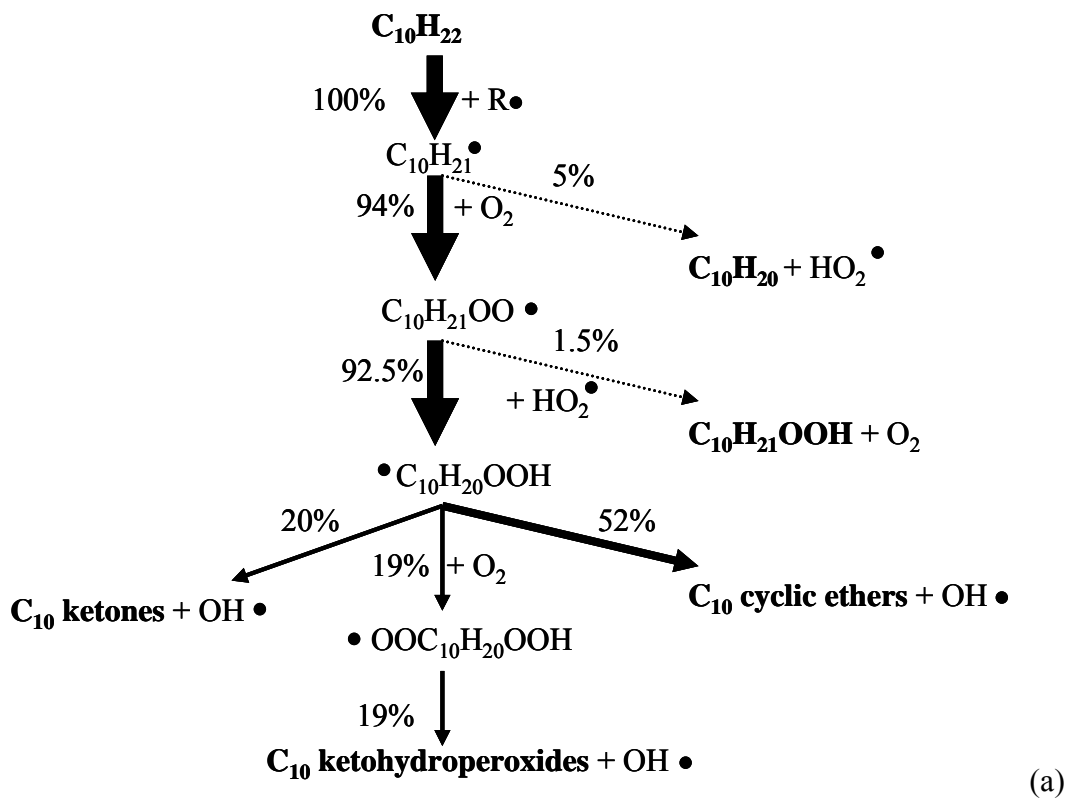

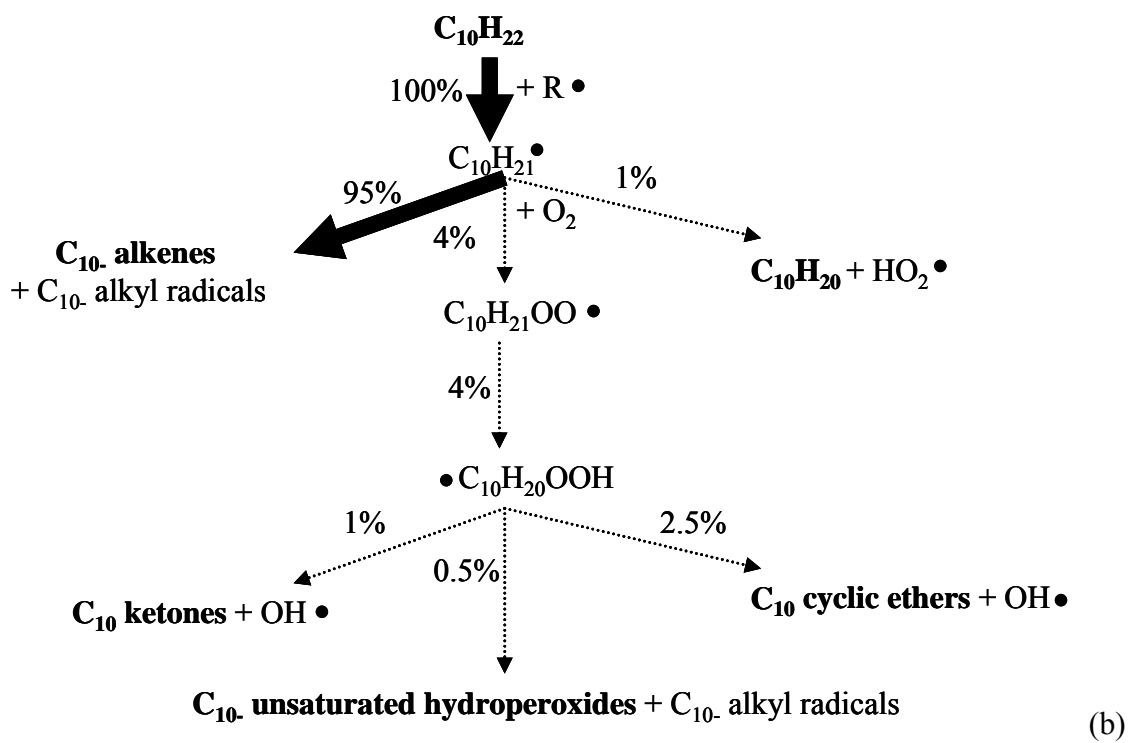

Figure 10: Flow rates analysis for the oxidation of n-decane in a jet-stirred reactor under the conditions of figure 2 at (a) 650 K and (b) 900 K.

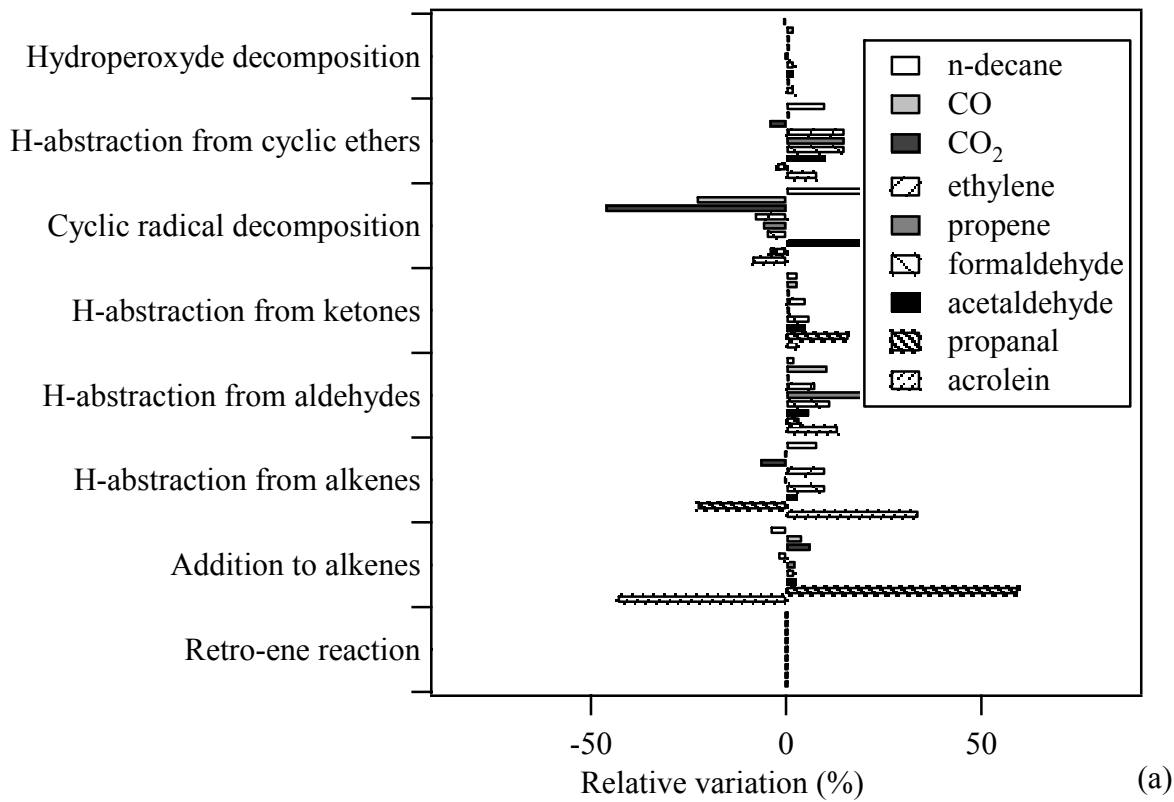

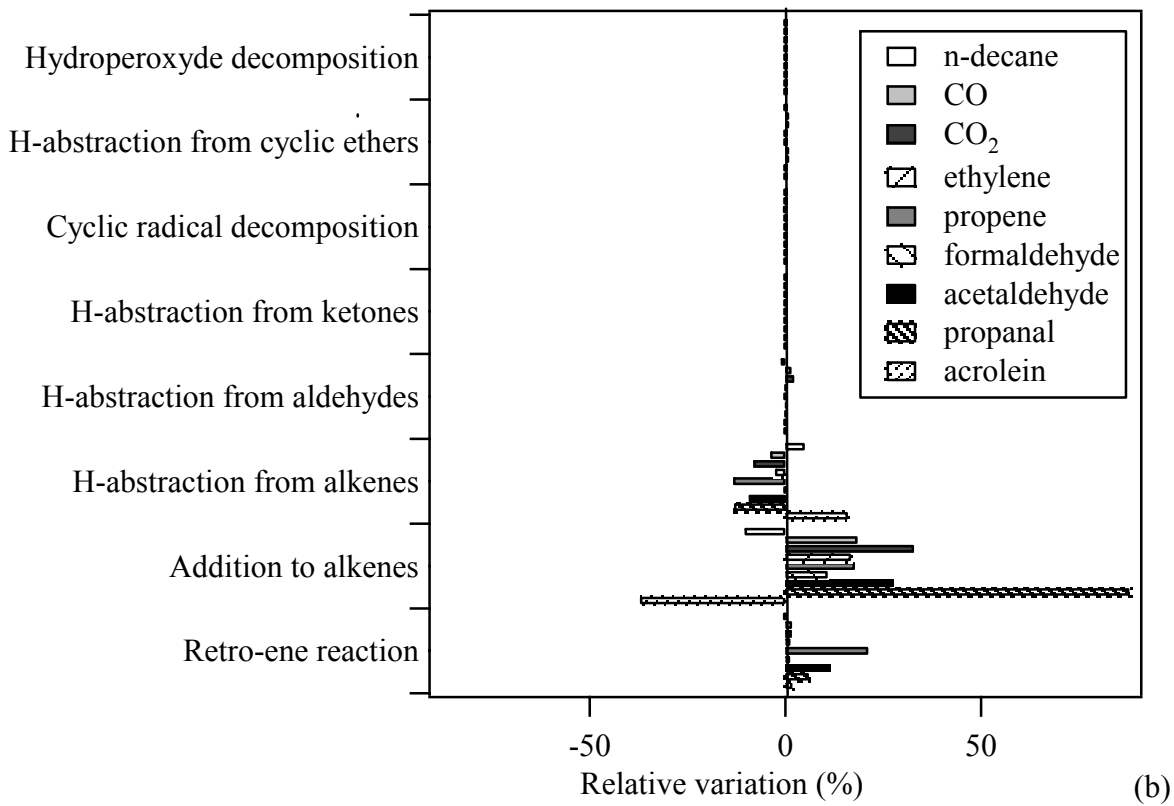

Figure 11: Sensitivity analysis for the reactions of primary products of $C_{3+}$ compounds during the oxidation of n-decane in a jet stirred reactor under the conditions of figure 2 at (a) 650 K and (b) 900 K.

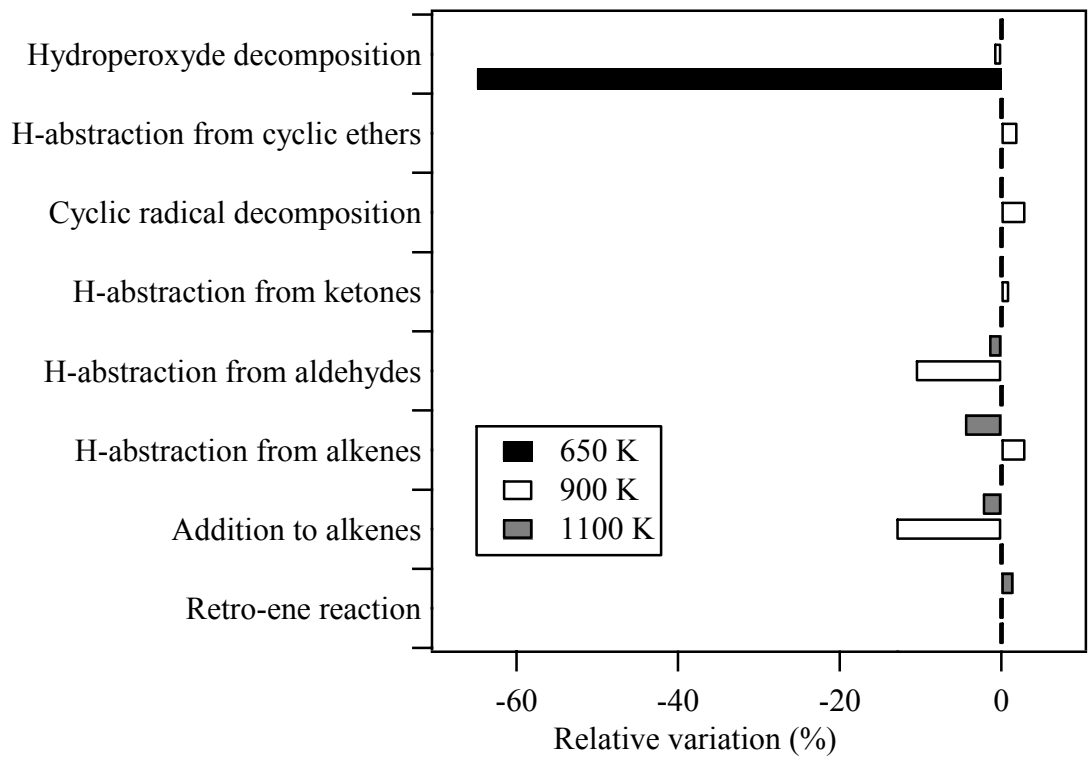

Figure 12: Sensitivity analysis for the reactions of primary products during the auto-ignition of n-decane in a shock tube under the conditions of figure 8a at 12 bar.

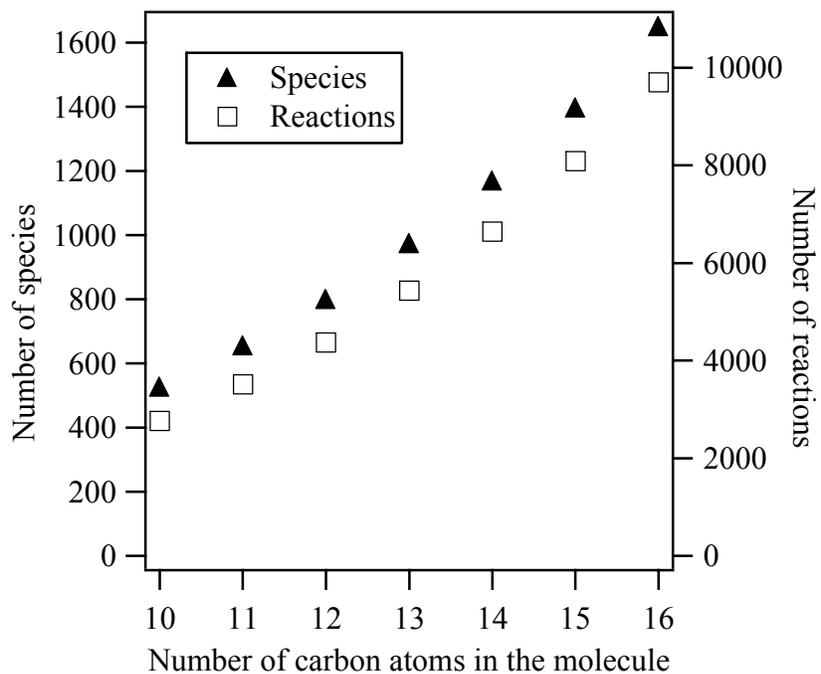

Figure 13: Variation of the number of species and reactions involved in the low-temperature oxidation mechanisms generated by EXGAS software with the number of carbon atoms in the reacting molecules for large alkanes.

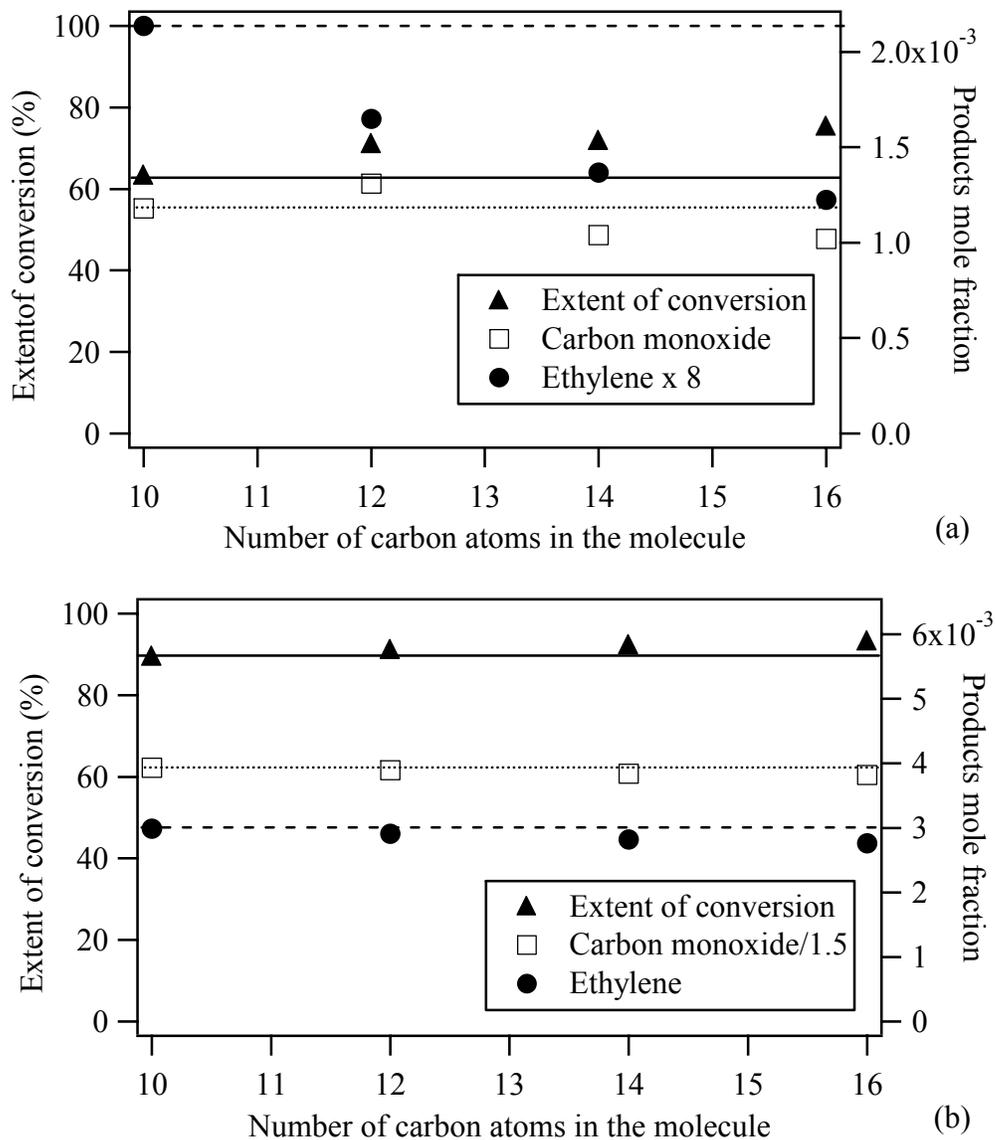

Figure 14: Variation of the simulated extent of conversion and mole fractions of carbon monoxide and ethylene with the number of carbon atoms in the reacting molecules during the oxidation of large alkanes in a jet-stirred reactor at (a) 700 and (b) 900 K (under the same conditions as figure 2 for n-decane and with the initial mole fraction of carbon atoms kept constant for other compounds). For each profile studied, the horizontal line plots the initial value.

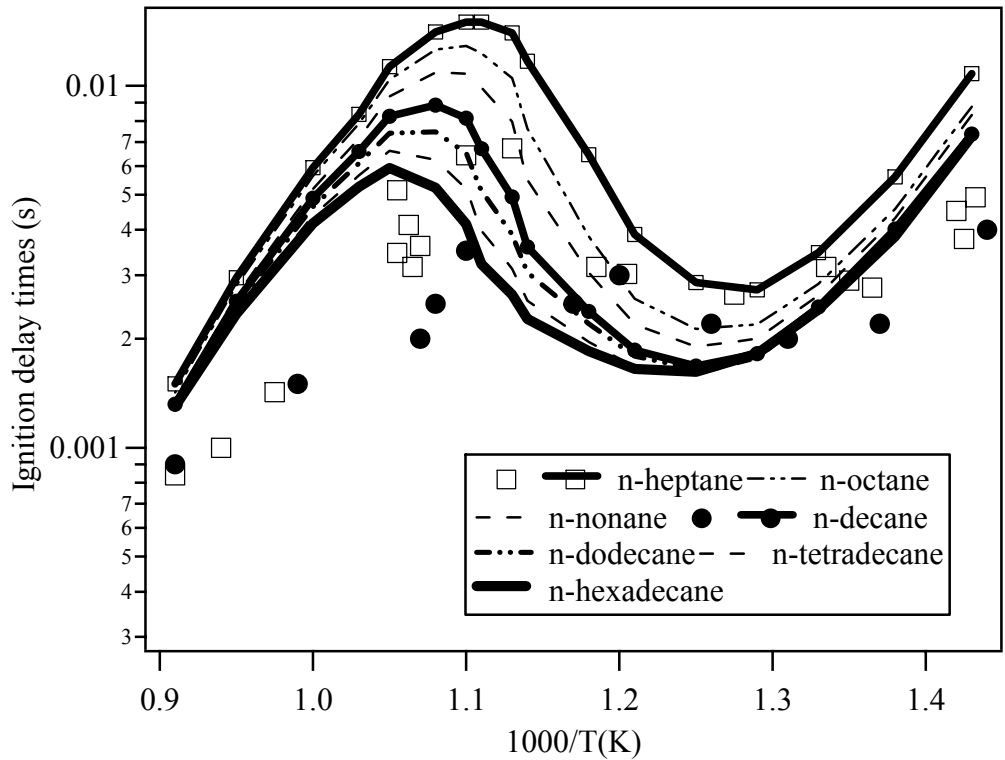

Figure 15: Computed ignition delay times for a series of linear alkanes for stoichiometric alkane/air mixtures at 12 bar. Symbols are experimental results from the team of Adomeit (11)(42) measured under these conditions.

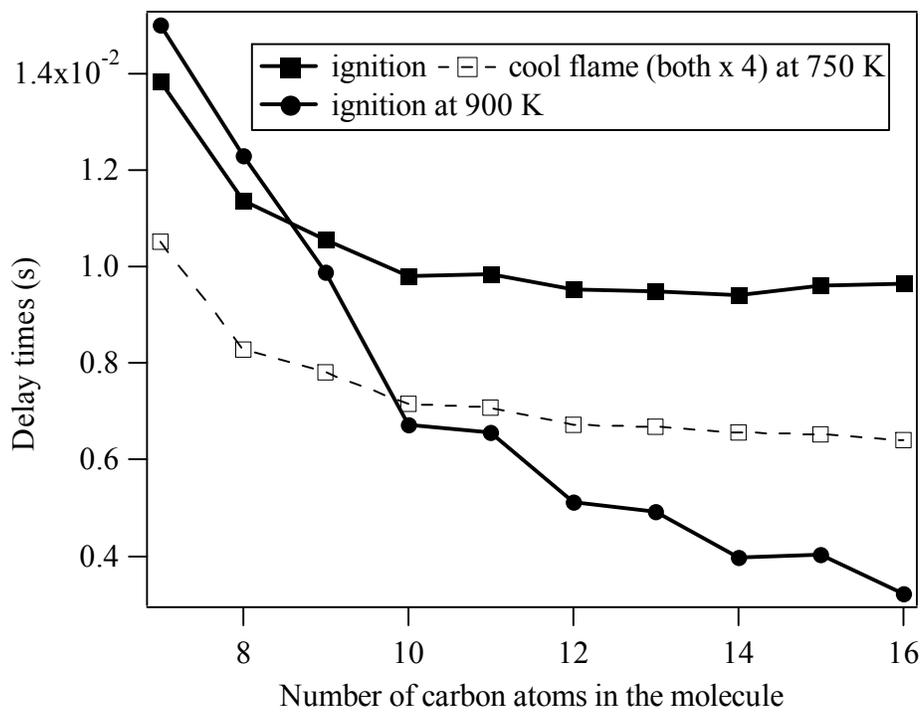

Figure 16: Variation of simulated cool flame and ignition delay times vs. the number of carbon atoms in the reactant for a series of linear alkanes for stoichiometric alkane/air mixtures at 12 bar.

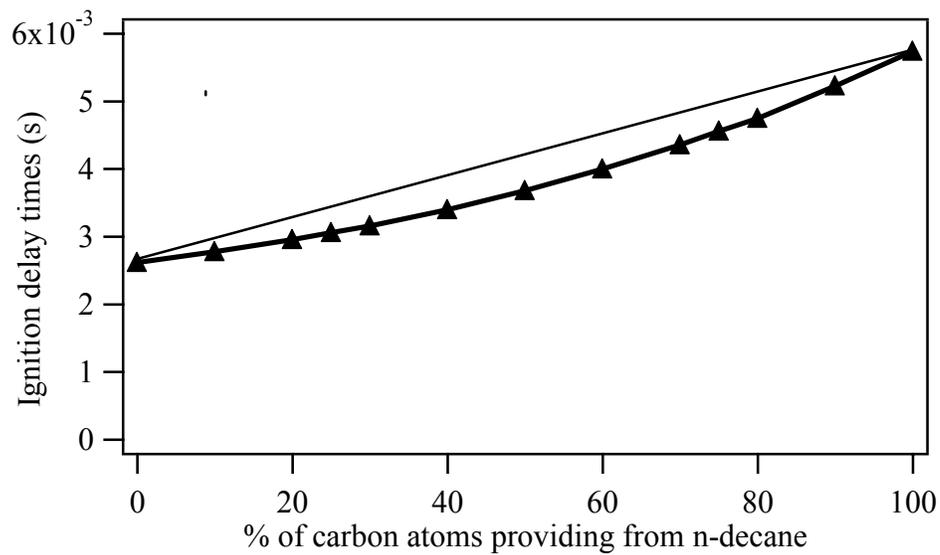

Figure 17: Variation of simulated delay times vs. the composition of the mixture (in % of carbon atoms providing by n-decane in the total number of carbon atoms) for n-decane/n-hexadecane blends for stoichiometric hydrocarbon/air mixtures at 12 bar and 900 K.